\documentclass[sigconf]{acmart}

\renewcommand\footnotetextcopyrightpermission[1]{} 
\usepackage{graphicx} 
\usepackage{caption}
\usepackage{subcaption}
\usepackage{bbm}
\usepackage{soul}
\usepackage[]{algorithm2e}

\def\BibTeX{{\rm B\kern-.05em{\sc i\kern-.025em b}\kern-.08emT\kern-.1667em\lower.7ex\hbox{E}\kern-.125emX}}
    
\copyrightyear{2019}
\acmYear{2019}
\acmConference[KDD '19 Applied Data Science Track]{}{August 04--08, 2019}{Anchorage, USA}
\acmBooktitle{KDD '19, August 04--08, 2019, Anchorage, USA}

\usepackage{bm}
\newcommand{\vk}{{\bm k}}

\newcommand{\vr}{{\bm r}}
\renewcommand{\d}{\mathrm{d}}
\newcommand{\pred}{\mathrm{pred}}
\newcommand{\true}{\mathrm{target}}

\begin{document}
\sloppy

%
\title[From Dark Matter to Galaxies with CNN]{From Dark Matter to Galaxies with Convolutional Networks}
%

\author{Xinyue Zhang*, Yanfang Wang*, Wei Zhang*, Yueqiu Sun*}
\authornote{The authors contributed equally to this work.}
\affiliation{%
  \institution{Center for Data Science, New York University}
}
\email{xz2139, yw1007, wz1218, ys3202@nyu.edu}

\author{Siyu He}
\affiliation{
\institution{Department of Physics, Carnegie Mellon University}
}
\affiliation{
\institution{Center for Computational Astrophysics, Flatiron Institute}
}
\email{she@flatironinstitute.org}

\author{Gabriella Contardo, Francisco Villaescusa-Navarro}
\affiliation{%
  \institution{Center for Computational Astrophysics, Flatiron Institute}
}
  \email{gcontardo, fvillaescusa@flatironinstitute.org}
 
 \author{Shirley Ho}
\affiliation{%
  \institution{Center for Computational Astrophysics, Flatiron Institute}
}
\affiliation{
    \institution{Department of Astrophysical Sciences, Princeton University}
}
\affiliation{
    \institution{Department of Physics, Carnegie Mellon University}
}
\email{shirleyho@flatironinstitute.org} 
  
%
\renewcommand{\shortauthors}{Zhang et al.}

%

\begin{abstract}

Cosmological surveys aim at answering fundamental questions about our Universe, including the nature of dark matter or the reason of unexpected accelerated expansion of the Universe.
In order to answer these questions, two important ingredients are needed: 1) data from observations and 2) a theoretical model that allows fast comparison between observation and theory. Most of the cosmological surveys observe galaxies, which are very difficult to model theoretically due to the complicated physics involved in their formation and evolution; modeling realistic galaxies over cosmological volumes requires running computationally expensive hydrodynamic simulations that can cost millions of CPU hours. In this paper, we propose to use deep learning to establish a mapping between the 3D galaxy distribution in hydrodynamic simulations and its underlying dark matter distribution. One of the major challenges in this pursuit is the very high sparsity in the predicted galaxy distribution. To this end, we develop a two-phase convolutional neural network architecture to generate fast galaxy catalogues, and compare our results against a standard cosmological technique. We find that our proposed approach either outperforms or is competitive with traditional cosmological techniques. Compared to the common methods used in cosmology, our approach also provides a nice trade-off between time-consumption (comparable to fastest benchmark in the literature) and the quality and accuracy of the predicted simulation. 
In combination with current and upcoming data from cosmological observations, our method has the potential to answer fundamental questions about our Universe with the highest accuracy. \footnote{The source code of our implementation is available at: \url{https://github.com/xz2139/From-Dark-Matter-to-Galaxies-with-Convolutional-Networks}}

\end{abstract}

%
%


%
%
\maketitle

\section{Introduction}
\label{submission}

Cosmology focuses on studying the origin and evolution of our Universe, from the Big Bang to today and its future. One of the holy grails of cosmology is to understand and define the physical rules and parameters that led to our actual Universe. Astronomers survey large volumes of the Universe \cite{6df, lsst,2dF, wfirst} 
and employ a large ensemble of computer simulations to compare with the observed data in order to extract the full information of our own Universe. 

The constant improvement of computational power has allowed cosmologists to pursue elucidating the fundamental parameters and laws of the Universe by relying on simulations as their theory predictions. These simulations can help determine if a set of rules or specific parameters can lead to the observed Universe. An important type of simulations  is gravo-hydrodynamical simulations, which aim at reproducing the formation and evolution of galaxies through time. 

However, evolving trillions of galaxies over billions of light years including the forces of gravity, electromagnetism, and hydrodynamics, is a daunting task. 
The state-of-art fully gravo-hydrodynamical cosmological simulations that include most of the relevant physics can only simulate a small fraction of our Universe and still requires 19 million CPU hours (or about 2000 years on one single CPU) for the most recent one \cite{illustris} to complete. 

On the other hand, the standard cosmological model provides us with a solution to this challenge: most of the matter in the Universe is made up of dark matter, and the large scale cosmic structure of the Universe can be modeled quite accurately when we evolve dark matter through time with only physics of gravity. When we do add the gas into the mix, gas usually traces the matter density, and for large enough dark matter halos, gas falls to the center of dark matter halos, subsequently cool down and form stars and galaxies. In other words, dark matter halos form the skeleton inside which galaxies form, evolve, and merge. Hence, the behaviors, such as growth, internal properties, and spatial distribution of galaxies, are likely to be closely connected to the behaviors of dark matter halos.

A gravity-only $N$-body simulation is the most popular and effective numerical method to predict the full 6D (position and velocity) phase-space distribution of a large number of massive particles, whose position and velocity evolve over time in the Universe \cite{Davis1985}. 
They are computationally significantly less expensive than when we include other complex physics such as hydrodynamics and astrophysical processes.  However, these simulations do not include `baryonic' information (i.e. galaxies distributions). To overcome this problem, different approaches have been proposed to map from the dark-matter distribution (obtained with gravity-only $N$-body simulations) to the galaxy distribution (see Section 2 for more details), but they suffer from a trade-off between the accuracy of important physical properties of the Universe's expected structure and the scaling abilities and time consumption. Besides, they usually rely on assumptions like halo mass being the main quantity controlling galaxy properties such as clustering.

We propose in this paper a first machine-learning based approach for this problem. We explore the use of convolutional neural networks (CNN) to perform the mapping from the 3D matter field in an N-body simulation to galaxies in a full hydrodynamic simulation. The task can be formulated as a supervised learning problem. One main difficulty of this application is the very high sparsity of the 3D output, compared to the input. We design to this end a specific two-step architecture and learning scheme that alleviates this problem. We evaluate our resulting using different statistics evaluated on the hydrodynamic simulation  (e.g. power-spectrum, bispectrum) to verify the accuracy of our predictions, and to compare with a benchmark method commonly used in cosmology. We show that our approach provides more accurate galaxy distribution than the benchmark on various criteria: positions, number of galaxies, power spectrum and bispectrum of galaxies.
 This illustrates a better fit of the different structures and physics properties one can evaluate on the galaxy distribution. Our method also benefits from great scaling ability: we could potentially generate large volumes of realistic galaxies in a very competitive time.

We provide background and review related works in cosmology and machine learning in Section 2. Section 3 presents the data used. Section 4 presents our model's architecture. We show quantitative results and visualizations of our predicted hydrodynamic simulation in Section 5. We conclude and discuss future works in Section 6.

\section{Background and Related Works}

\subsection{Cosmology} 
 
The 21st century has brought us tools and methods to observe and analyze the Universe in far greater detail than before, allowing us to probe the fundamental properties of cosmology. We have a suite of cosmological observations that allow us to make serious inroads to the understanding of our own Universe, including the cosmic microwave background (CMB) \cite{planck18, 2013ApJS..208...20B} supernova \cite{perlmutter99} 
and  the large scale structure of galaxies \cite{cole05, anderson14}. In particular, large scale structure involves measuring the positions and other properties of bright sources in great volumes of the sky. These observations established a best model of the Universe, which is currently described by
less than 10 parameters in the standard $\Lambda {\rm CDM}$ cosmology model, where CDM stands for cold dark matter and $\Lambda$
stands for the cosmological constant. The parameters that are important for this analysis include the matter
density $\Omega_m \approx 0.3$ (normal matter and dark matter together constitute approximately 30\% of the energy content of the Universe), the variance in the matter overdensities $\sigma_8 \approx 0.8$ (the variance of the matter field density on spheres of 8 Mpc/h), and the current Hubble parameter $H_0 = 100h $ $\approx  70$km/s/Mpc (which describes the present rate of expansion of the Universe). The model also assumes 
a flat geometry for the Universe.
Note that the unit of distance megaparsec/h ( Mpc/h )
used above is time-dependent, where 1 Mpc is equivalent
to $3.26 \times 10^6$ light years and h is the dimensionless Hubble
parameter that accounts for the expansion of the Universe.

The amount of information collected by modern astronomical surveys is overwhelming, and modern
methods in machine learning and statistics can play an increasingly important role in modern cosmology. For example, the traditional method to compare large scale structure observation and theory relies on matching the compressed
two-point correlation function of the observation with the theoretical prediction (which is very hard to model on small scales, where a significant cosmological information lies). There are several other methods in the cosmological community that allow cosmologists to utilize more information from the astronomical surveys and here we discuss some of them. 

\textbf{Mocking up the Universe with Halo Occupation Distribution Model} 
Halo Occupation Distribution model\cite{2007MNRAS.374..477T, 2002ApJ...575..587B, vakili2016galaxies} (hereafter HOD) is widely used to connect dark matter halos and galaxies. 
HOD is based on the assumption that the probability of the presence of a galaxy at certain position on the sky (or simulation) is based solely on the mass of the dark matter halo that the galaxy will sit in. 
The average number of galaxies in a halo of a certain mass is a function of only the halo mass. It describes how the distribution of the galaxies is related to the distribution of the dark matter halos, therefore providing us with 
a way to populate N-body simulation of dark matter particles with galaxies, allowing us a direct way to compare observed distribution of galaxies on the sky and our theoretical predictions represented by simulations. However, this method comes with its own limitation: multiple tuning parameters, all galaxies live in dark matter halos 
and the assumption that halo mass in the main property controlling the abundance and clustering of galaxies. 
 
\textbf{Mocking up Universe with Abundance Matching} 
Abundance matching is a popular method to connect dark matter halos with galaxies, by ranking the dark matter halos by mass and the galaxies by luminosities. 
We then match brighter galaxies to the heavier halos, and we keep going down the ranked lists. Similar to the HOD, this is an easy way to populate N-body simulations of dark matter particles with galaxies, allowing direct 
comparisons between observed distribution of galaxies in the sky and our theoretical predictions represented by the simulations. This method also relies on assumptions, like monotonic relations between halo mass and galaxy abundances.

\subsection{Machine Learning in Cosmology} 

Convolutional neural networks are traditionally used in computer vision tasks, such as image classification, detection, and segmentation. They are increasingly being adopted in cosmology researches nowadays, and work well in representing features of Universe. 
\cite{ravanbakhsh2016} estimates cosmology parameters from the volumetric representation of dark-matter simulations using 3D convolutional networks with high accuracy. They showed that machine learning techniques are comparable to, and can sometimes outperform cosmology models. The paper identifies ReLU, average pooling, batch normalization and dropout as critical design choices in the neural network architecture to achieve highly competitive performance in estimating cosmology parameters. 
\cite{kamdar2016machine} used Extremely randomized Trees \cite{Geurts2006} to predict a hydrodynamical simulation of galaxies and found that ERT is very efficient in reproducing the statistical properties of galaxies in these hydrodynamical simulations.  
In \cite{ribli2018learning}, CNN has been demonstrated to give significantly better estimates of $\omega_m$ and $\sigma_8$ cosmological parameters from simulated convergence maps than the results from state-of-art methods, but is also free of systematic bias. Additionally, the CNN model could be interpreted by using the representations from internal layers. The similarity between a kernel and the Laplace operator inspired Ribli et al. to propose a new peak counting scheme that achieves better result than past peak counting schemes. %

\section{Data}
\label{sec:data}

\begin{figure}[t]
    \centering
    \includegraphics[width= 0.95\linewidth]{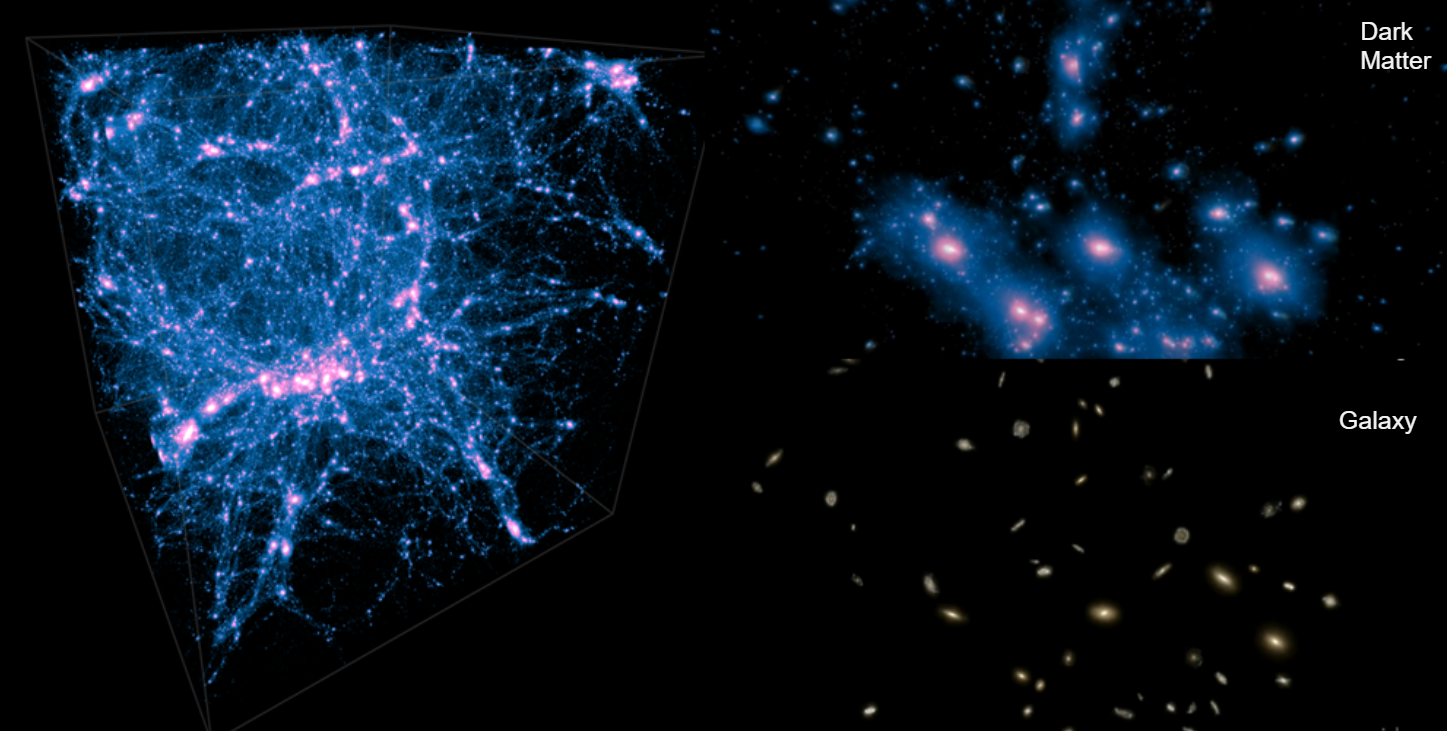}
    \caption{Visualization of Illustris simulation at redshift $z=0$ (left), and Zoom-in visualization of corresponding dark matters and galaxies. (Right), adapted from \url{http://www.illustris-project.org/media/}
}
\label{fig:snapshotIllustris}
\end{figure}

We utilize two types of simulations in this work: hydrodynamic (gravitational + hydrodynamic forces and astrophysical processes) and N-body (only gravitational forces), both from the Illustris project \cite{illustris,illustris_release,illustris_data,illustris_galaxy}. The main purpose of this work is to train neural networks to predict the abundance and spatial distribution of galaxies from the very computationally expensive hydrodynamic simulation only using information from the much cheaper N-body simulation.

We use the level-1 simulations within Illustris, which is the simulation set with highest spatial and mass resolution of the suite. 
We focus our analysis at redshift $z=0$, that corresponds with the current epoch of the Universe.

The cosmological model used for the Illustris simulation is in agreement with the constraints from WMAP9 (Nine-Year Wilkinson Microwave Anisotropy Probe Observations)\cite{WMAP9}.

At $z=0$, the hydrodynamic simulation contains 5,280,615,062 gas cells, 595,243,070 stellar particles and 32,552 supermassive black-holes particles. The number of Dark Matter particles and galaxies within this snapshot is 6,028,568,000 and 4,366,546, respectively. The N-body simulation only contains 6,028,568,000 dark matter particles. Figure \ref{fig:snapshotIllustris} shows the spatial distribution of dark matter in the N-body simulation as well as a close-up on a smaller region.

We compute the density fields of galaxies and dark matter by assigning each component to a regular grid with $1024^3$ voxels using the nearest-grid point mass assignment scheme: if a galaxy or dark matter particle is inside a given voxel, the value of that cell is increased by 1 for the corresponding field.
 
\begin{figure}[t]
\begin{center}
    \includegraphics[width=\linewidth]{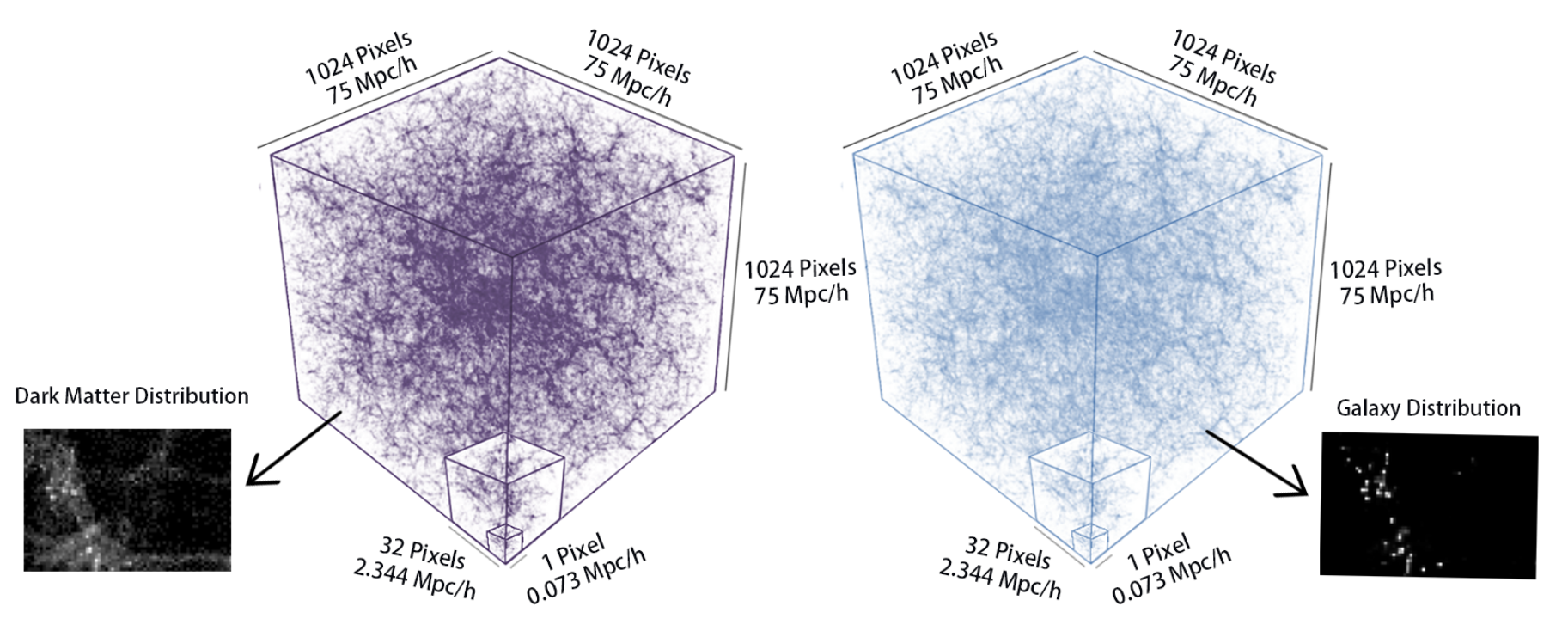}
    \caption{Spatial distribution of dark matter from the N-body simulation (left) and galaxies from the hydrodynamic simulation (right). The large boxes represent the entire simulations, while the small boxes correspond to the siezes of our voxels and training cubes.
    }
    \label{illustris_v}
\end{center}
\end{figure}

Within the grid, the number of particles in each voxel ranges from 0 to 747,865 for dark matter and from 0 to 10 for galaxies. The percentage of non-zero cells is 44.99\% and 0.37\% for dark matter and galaxies, correspondingly. The low occupancy of galaxies in the grid poses an interesting challenge for our work. 
Fig. \ref{illustris_v} shows the distribution of dark matter and galaxies from the N-body and hydrodynamic simulations, respectively. The voxels are shown to demonstrate the gridding we perform. 

The simulation density fields are then separated into sub-cubes of size $32^3$ voxels, corresponding to regions of size around 2.3 Mpc/h, and are used as independent samples. There are 32,768 unique sub-boxes and they are split spatially into three chunks. 62.6\% of all boxes are used for training, 19.63\% of the boxes are used for validation and then the other 17.76\% are used for testing. Testing data are retained as a concatenated cube of size 42.4 Mpc/h for the ease of computation of relevant statistics.

\section{Methods}
\label{sec:methods}

Here we present our approach for linking the 3D dark matter field from N-body simulations to the 3D galaxy distribution from hydrodynamic simulations.

The two key challenges in predicting galaxy positions from the dark matter field are (i) the inherently spatial nature of the data (dark matter and galaxies are structured spatially, on various correlated scales), (ii) the high sparsity of the galaxy (target) distribution.

To address the first aspect, we propose to rely on convolutional networks. They naturally provide interesting properties for our problem such as translational invariance \cite{2016RSPTA.37450203M}. Convolutional networks are also commonly employed for extraction of spatial patterns \cite{ravanbakhsh2016}. To address the second aspect, we developed a \textit{two-phase} architecture and learning process. We present in the following section the details of this architecture, and we discuss the different convolutional networks we tested in our experiments.

\subsection{Two-phase architecture}
The high sparsity in our simulation dataset (99.6$\%$ of output voxels do not contain any galaxies) makes our training challenging. Because of imbalanced distribution between input and output, the model could easily achieve a high accuracy even if it fails in predicting all the galaxies. This slows down the training process to a great extent. In order to overcome this problem, we propose the following two-phase architecture.

The main idea is to break down the training into separate processes. The whole model is composed of two parts. The first part is a classifier, which predicts the presence or absence of galaxies as a probability for each voxel representing one part of Universe. Using a binary classifier as a first "layer" allows us to use special loss functions designed for such high sparsity prediction. Specifically, we use weighted cross-entropy loss, which penalizes wrong predictions with high probability, and introduces weights to correct imbalances in classes. For a single output voxel, it can be written as follows:
~\\~\\
\begin{eqnarray}
\mathbb{L}_{\mathrm{CrossEnt}}(\hat{p},\mathbf{y})=-( w \cdot \mathbf{y} \cdot \mathrm{log}(\hat{p}) + (1-\mathbf{y})\cdot \mathrm{log}(1 - \hat{p}))~
\end{eqnarray}
 
where $\hat{p}$ is the vector of predicted probability of the presence of at least one galaxy in the considered voxel. $\mathbf{y}$ is the actual target value (1 if there exists at least 1 galaxy in the voxel, 0 otherwise). $w$ characterizes the weight applied for counter-balancing the large number of voxels without galaxies\footnote{Details on the selection for $w$ are given in Section \ref{sec:results} and in Supplementary Materials}.

This first prediction is then used as a mask for the final prediction of the number of galaxies in each voxel. The second step of the network is optimized only on the voxels that are expected to contain at least a galaxy, according to the binary prediction result from the first phase. We propose to use a $L2$-loss since we decide to predict a probabilistic number of galaxies in each voxel, and expect to have a real value as output. The complete loss of the model for a single output voxel is illustrated below: 
~\\~\\
\begin{eqnarray}
\mathbb{L}(n_g,\hat{p},n_t) = M(\hat{p})(n_g - n_t)^2
\end{eqnarray}

\begin{eqnarray}
  M(\hat{p})=\begin{cases}
    1       \quad \hat{p} > 0.5\\
    0       \quad Otherwise \\
            \end{cases}
\end{eqnarray}

Where $M$ is a function of the first-phase output ($\hat{p}$) for the given voxel, which returns 1(0) if we expect to see at least 1 galaxy in the voxel (or not).  $n_g$ is the prediction of the second-phase model and $n_t$ is the actual target: number of galaxies in the output voxel. 

\begin{figure}
\begin{center}
    \includegraphics[width= \linewidth]{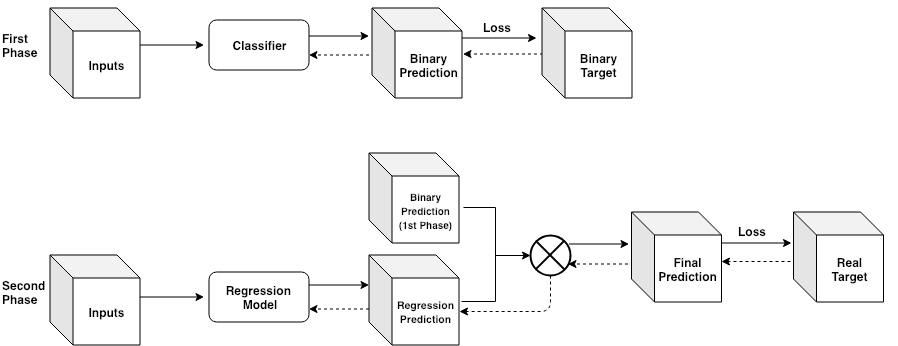}
    \caption{Two Phase Model Structure }
    \label{fig:twophase}
\end{center}
\end{figure}

From an experimental point of view, the training of both parts is done separately. We select the first classifier (e.g. with the highest recall) and then train the second part of the model. More details are given in Sec~\ref{sec:results}. A schema of this generic architecture is shown in Figure \ref{fig:twophase} for a better visualization of the process.

This two-phase set-up is quite generic, and allows us to build different types of architecture depending on the choice of networks for each phase.

\subsection{Network architectures}
We now present different types of convolutional networks we tested, with their physical motivations and specific modifications to better fit our problem. The models mentioned below can be used for both classification(first-phase) and regression(second-phase).  We will compare different choices of networks in Section ~\ref{sec:results}, as well as a more classical "one-phase" training.

\paragraph{U-Net} 

U-Net is a fully convolutional neural network, first proposed in \cite{ronneberger2015u} for bio-medical image segmentation. The networks is composed of a \textit{contracting path} and a symmetric \textit{expanding path}. The first path is a typical convolutional architecture, with convolutions followed by a rectified linear unit (ReLU) and max-pooling operation. The number of channels is increased at each step. This part aims at capturing spatial relations and context. The expanding path relies on up-sampling functions on the feature map, followed by \textit{up-convolutions} that reduce the number of channels. Additionally, a \textit{skip connection} is added at each level, which concatenates the up-sampled features and the corresponding map from the contracting path. This part provides the network with various levels of granularity for the final prediction, usually segmentation, which is in a similar shape as the input.

This type of network can be easily adapted to 3D data and has been successfully applied in different applications, for instance in cosmology \cite{he2018learning}, or for volumetric segmentation on medical data \cite{cciccek20163d}. As its architecture is constructed to map between input and output with similar shapes and to extract spatial information on multiple scales, it appears as a good candidate to learn the relationship between dark-matter halos and the distribution of galaxies. 

\begin{figure}
\begin{center}
    \includegraphics[width=\linewidth]{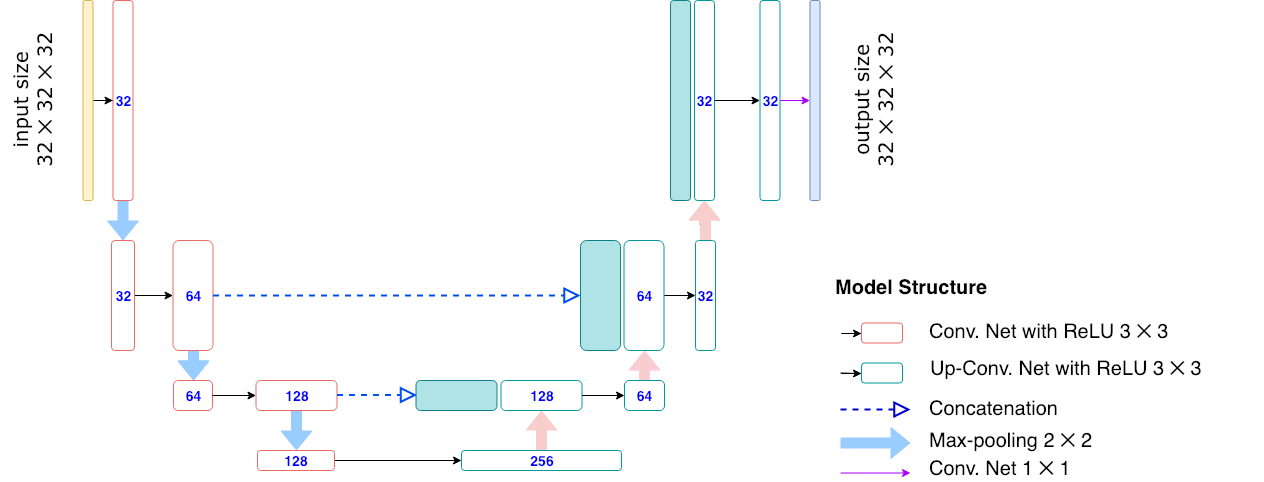}
    \caption{U-Net architecture with removed \textit{skip}-connection on the upper layer.}
    \label{Unet}
\end{center}
\end{figure}

Our early experiments on this model structure showed that the prediction had a strong similarity in distribution with inputs instead of targets, which constrains the model from generalizing to larger scale. Thus, we proposed one modification to the original architecture. The last (topmost) skip-connection was removed to prevent the model from "feeding" too much information from the dark-matter halos on high resolution features. The final U-Net architecture used in the experiments is illustrated in Figure \ref{Unet}.

\paragraph{Recurrent Residual U-Net (R2Unet)}
Recurrent Residual U-Net (R2U-Net) were proposed in \cite{MdZahangR2Unet} as an upgrade of U-Net. The authors propose different variations around the U-Net architecture, but we focus here on the Recurrent-Residual one. The main idea is to change the convolution functions used in the U-Net architecture. Instead of using classical convolution functions, R2U-Net relies on a composition of two stacked Recurrent Convolutions (RCNN), as presented in \cite{pinheiro2014recurrent}, with a residual connection from the input to the output. The recurrent aspect of RCNN allows to produce arbitrarily deeper architecture without increasing the number of parameters, while allowing the model to refine and aggregate the extraction of features through "time" (steps applied during the recurrent convolution). This helps to accumulate the feature representation. The residual connection on the other hand allows for the propagation of higher level information at each layer.

We propose to use this model, here again with the modified U-Net global architecture where we remove the upper skip connection.

\begin{figure}
\begin{center}
     \includegraphics[width=\linewidth]{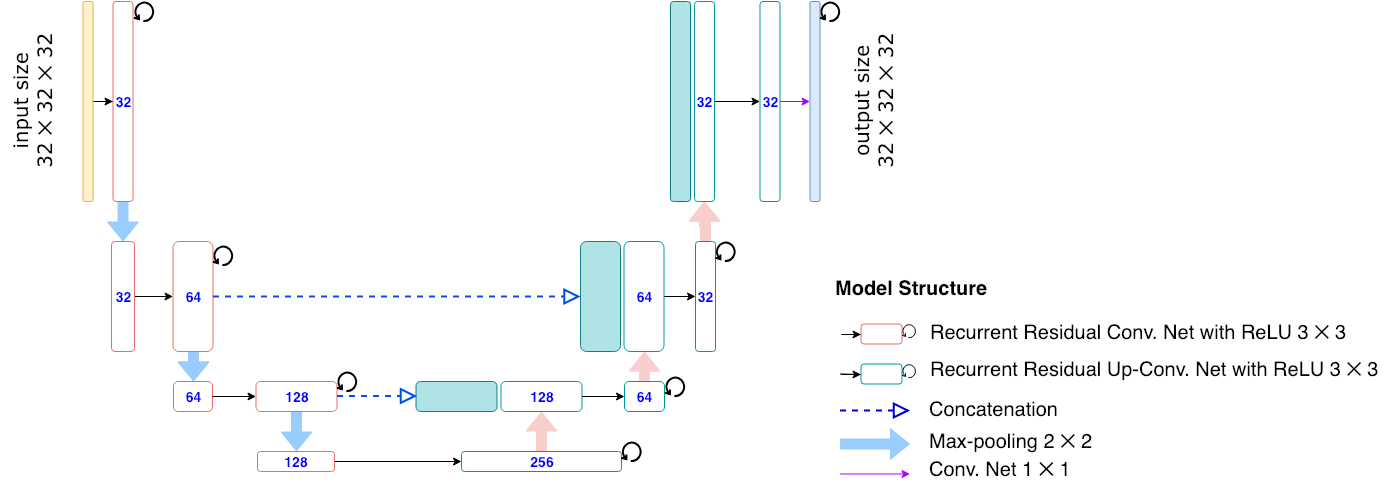}
    \caption{Recurrent Residual U-Net (R2U-Net) with removed upper skip-connection.}
    \label{R2UNET}
\end{center}
\end{figure}

\paragraph{Inception Net}

Inception networks \cite{szegedy2015going} were developed to handle the variation at the scale of the salient parts of images. 
The salient information can come in multiple scales . With "vanilla" convolutional networks, this leads to the difficulty of choosing the kernel size for the convolution function: if the information is structured on larger scales, one should choose a bigger kernel size, and inversely if the information is more locally distributed. Inception module proposes to horizontally stack convolution functions with different kernel sizes. Different variations around this key idea have been proposed in order to optimize the computation cost of widening the architecture. 

\begin{figure}
\centering
\begin{minipage}{0.5\linewidth}
  \centering
    \includegraphics[width= \linewidth]{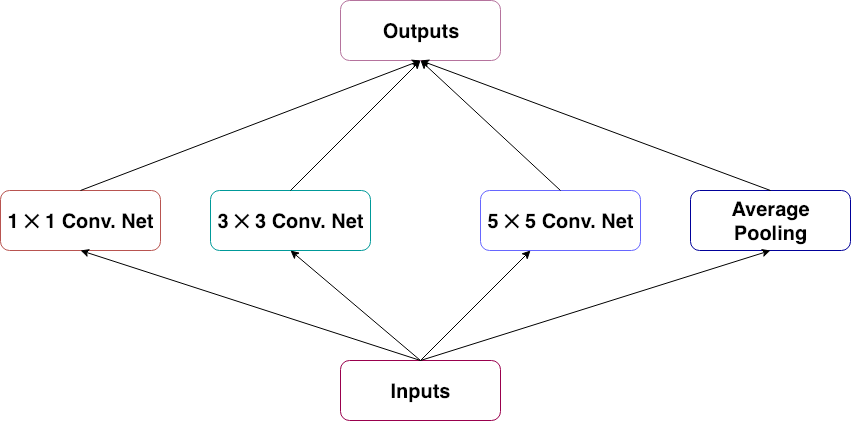}
    \caption{Inception Module v1 used in our network.}
    \label{incep3}
\end{minipage}%
\hspace{0.2cm}
\begin{minipage}{0.45\linewidth}
  \centering
    \includegraphics[width=0.19\linewidth]{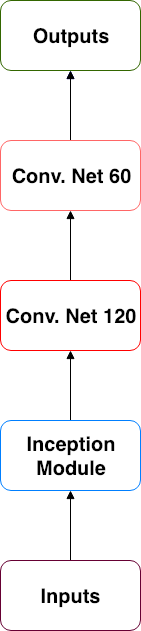}
    \caption{Architecture of the neural network used with Inception module (called in the remaining of the paper \textit{Inception}).}
    \label{incep4}
\end{minipage}
\end{figure}

In our application, it is very likely that the information extracted from the dark-matter halos is distributed on different scales, globally and locally. This motivates the use of the Inception module. More specifically, we propose to use the first original version of the Inception architecture module with 3 different kernel sizes of the 3 dimensional filters (1x1x1, 3x3x3, 5x5x5) and pooling layer. We use average pooling as we found it yields better results. The outputs of all the filters are concatenated and passed on as input to the subsequent layers. To limit the amount of parameters, we use convolutional functions as subsequent layers instead of fully connected ones, more specifically two convolutional layers to match the size of the target output. We use Sigmoid function as the final activation function of CNN network to output the probability that there is at least a galaxy in the voxel. Figure \ref{incep4} shows the whole model structure.

\section{Results}
\label{sec:results}

In this section we describe the results we obtain with different architectures of convolutional neural networks and compare them with a benchmark method commonly used in cosmology, HOD (Halo Occupation Distribution). Our main goal is to accurately predict the abundance and spatial distribution of galaxies by using information only from the 3D dark-matter field. By using different cosmological observables, we show how our method outperforms, or is competitive with HOD on all the considered observables.

\subsection{Experimental protocol}

As described in Sec~\ref{sec:data}, our models take as input the density field of dark matter from the N-body simulation in 3D sub-boxes containing $2^{15}$ voxels, and predict the 3D galaxy field from the hydrodynamic simulation. We retain 62.6\% of the $2^{15}$ total sub-boxes for training, 19.63\% for validation and 17.76\% for test. The split between training, validation, and test is made following a "global" cut, more specifically to enforce that the test sub-boxes form a larger cube with  $42.2$ Mpc/h on the side. This is motivated by the desire to compare our observables in a larger range of scales. 
\\~\\
We compare our results to those from \textit{halo occupation distribution} (HOD) algorithm (see e.g. \cite{berlind2002halo, villaescusa2014cosmology}), a method commonly used to link dark matter halos to galaxies in cosmology. The underlying idea behind the HOD is that all galaxies reside within halos, and galaxies can be split into centrals and satellites. Our HOD has three free-parameters: $M_{\rm min}$, $M_1$ and $\alpha$ and the algorithm is as follows. Only halos with masses greater than $M_{\rm min}$ will host a central galaxy, that will be placed on the halo center. The number of satellites galaxies follow a Poisson distribution with mean $(M/M_1)^\alpha$, that are placed randomly within the dark matter halo.
 
Given a set of ($M_1$, $\alpha$), we fit for $M_{min}$ to match the predicted galaxy density with the true galaxy density from the target simulation with a threshold of 0.001. The only free-parameters are thus $M_1$ and $\alpha$. They are optimized by minimizing the squared difference between the power spectrum computed on predicted galaxies and the power spectrum observed on the actual galaxies. In the following experiments, these parameters are optimized on the test sub-box observations.

\subsection{Galaxy distribution - binary prediction and quantity prediction}

Trying to optimize directly the number of galaxies per voxel proved ineffective due to the high sparsity of the output data. 
This motivates us to first perform a binary prediction, which predicts whether there is at least one galaxy in a voxel.  
We use accuracy, recall and prediction as metrics to evaluate the performance of different models. However, because of the sparsity of the data, high accuracy doesn't necessarily represent a good model as a model that predicts every voxel to have zero galaxies will achieve an accuracy of 99.57\%. Table \ref{ta:weight} shows accuracy, recall and prediction for different models.
Our experiments show that the Inception-based network provides the best recall at 95.72\%. We observe that in this binary-setup, HOD also performs well in terms of trade-off between recall and precision, with a high accuracy.

Following the results of binary prediction, we select the model with the highest recall as our first-phase model. By doing so, we alleviate the sparsity problem in the data, which allows for a small unrecoverable error in accuracy but manages to obtain a higher precision for the prediction in the second phase model. The second-phase model therefore focuses on reducing number of false positives (aka. improving precision) and predicting the a probabilistic number of galaxies in each voxel. 
Table \ref{ta:mse} shows mean square error for different machine learning models and our HOD benchmark. The model that yields best performance is the two phase model with Inception network as first phase, and R2U-Net as second phase. Our approach significantly outperforms HOD in this set-up, which seems to indicate that while HOD predicts correctly the region of absence/presence of galaxies, it is much more imprecise when predicting the number of galaxies in each voxel. We will see in the next subsections that this will impact the different statistical measures one can make on the Universe.

\begin{table}
\centering
\caption{Performance on binary prediction of galaxies}

\begin{tabular}{p{2cm}llll}
\toprule  
Model                        & Configuration            & Accuracy & Recall & Precision \\ \midrule

Inception & Weight: 80          & 96.32   & \textbf{95.72} & 10.15 \\
U-Net        & Weight: 5           & 99.6   & 59.8 & 39.2    \\
R2Unet                       & Weight: 5           & 99.52    & 63.17  & 41.91     \\
R2Unet                       & Weight: 10          & 99.29    & 74.8   & 32.42     \\
R2Unet                       & Weight: 25          & 98.8     & 84.31  & 21.05     \\
R2Unet                       & Weight: 80          & 97.41     & 92.49  & 13.52     \\
HOD                    &     ---      &    \textbf{99.93}   & 86.80 & \textbf{94.76}   \\

\bottomrule
\end{tabular}
\label{ta:weight}
\end{table}

\begin{table}
\centering
\caption{Mean-Square Error evaluation for number of galaxies prediction}
\begin{tabular}{p{2cm}llll}
\toprule  
Model  & Configuration & MSE \\ \midrule
R2Unet+R2Unet &Weight: 80/0.6  &  0.00320 \\
Inception+R2Unet & Weight: 80/0.8 & \textbf{0.00308} \\
HOD & --- &0.01007\\
\bottomrule
\end{tabular}
\label{ta:mse}
\end{table}

\begin{figure}
\begin{center}
    \includegraphics[width=0.95\linewidth]{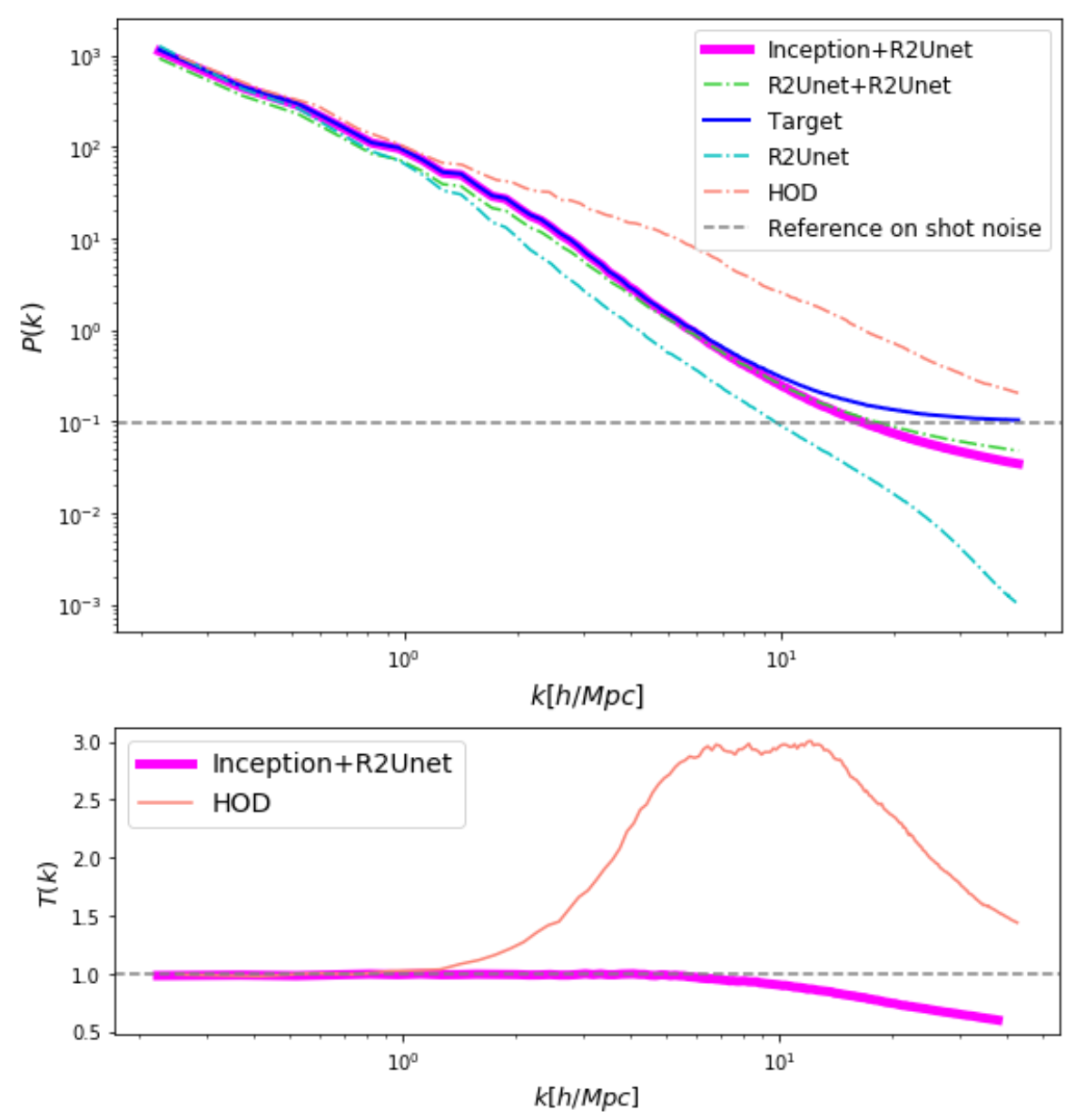}
    \caption{Power spectrum (Top) and Transfer Function (Bottom) comparison among different machine learning models, HOD model and target. The reference shot noise is the Poisson noise in the power spectrum resulting from Poisson sampling. Two-phase models performs as good as HOD on large scales (left of the upper plot, k<1 $h$/Mpc) and outperforms HOD on smaller scales.}
    \label{fig:power}
\end{center}
\end{figure}

\begin{figure*}
\hspace{-1cm}
\begin{center}
    \vspace{-1cm}
    \includegraphics[width=0.97 \linewidth]{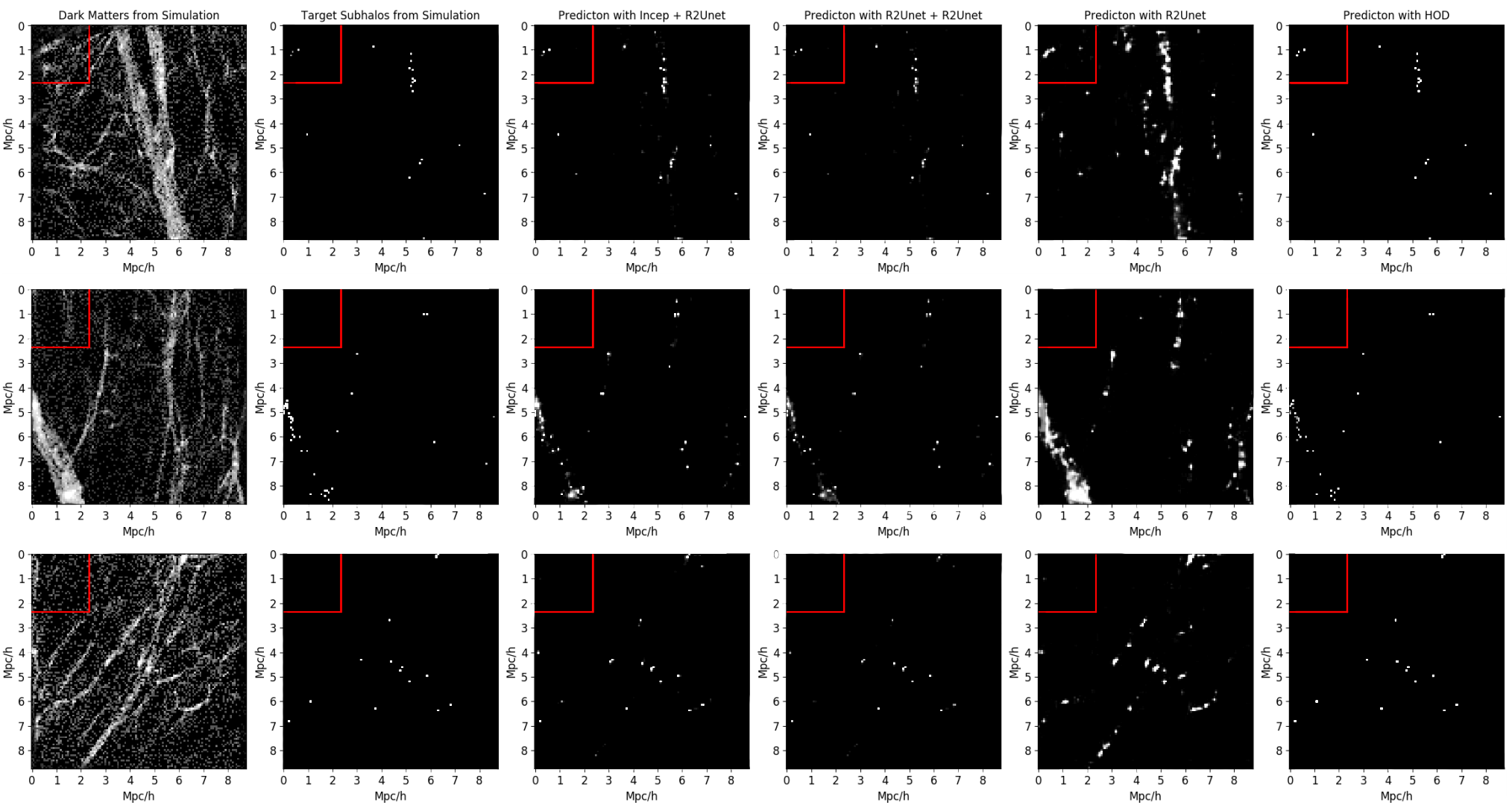}
    \caption{Visualization of slices of the simulations: first column are dark-matter halos, second column are the corresponding target galaxies. 3d and 4th columns are predictions from our two-phase models, 5th from a single-phase classifier, and last column are HOD predictions. Red square represents the size of the boxes taken as input by our models.}
    \label{fig:sampleresult}
\end{center}
\end{figure*}

A visualization of the predictions of different models is provided in Figure \ref{fig:sampleresult}. Each row represents a "slice" of the simulations, with $8.89$ Mpc/h on the side. As a reminder, the sub-boxes taken as inputs and outputs of our models are of size $2.34\,{\rm (Mpc/h)}^3$, and are depicted as a red square on the upper-left-most of the Figures. The left-most column is the dark-matter halos' masses taken as input. The second column depicts the corresponding target number of galaxies, where brighter pixels represent a higher number of galaxies. We can directly observe the high sparsity of the problem. The two-phase models predictions are depicted in the third and fourth column. In fifth column are predictions from a single-phase model. This visualization also illustrates the behavior of a single-phase training, which has trouble refining the galaxies predictions. Last column are HOD predictions. 

\subsection{Two-Point Correlation and Power Spectrum}

It is a standard practice in cosmology to extract information from observations via summary statistics. The most commonly used statistics is the two-point correlation function $\xi(r)$, defined as the excess probability, compared with that expected for a random distribution, of finding a pair of galaxies at a separation. It measures how the actual distribution of galaxies deviates from a simple random distribution. The power spectrum, $P(k)$, is the Fourier transform of the two-point correlation function:

\begin{equation}
    \begin{split}
    \xi(|\vr|) &= \langle \delta_A(\vr')\delta_B(\vr'+\vr) \rangle \\
    P(|\vk|) &= \int
    \!\d^3 \vr\;
    \xi(r)e^{i\vk \cdot \vr}
    \end{split}
\end{equation}

These two statistics are very important in cosmology, because they allow to extract all the information embedded into Gaussian density fields, as our Universe resembles on large-scales or at earlier times. In this paper we focus our attention on the power spectrum. We define the transfer function, $T(k)$, as
\begin{equation}
\label{transfer}
T(k) = \sqrt{\frac{P_\pred(k)}{P_\true(k)}}
\end{equation}
and use it to quantify the performance of the models against the ground truth.

Figure.~\ref{fig:power} shows the power spectrum and transfer function for the different models. Our two-phase model with Inception+R2Unet manages to reproduce the clustering of galaxies of the original data. Interestingly, it manages to obtain a good fit on a large range of scales, even though it is trained on relatively "small" sub-boxes. Comparing to the HOD results, our model achieves nearly the same performance when $k<1$ h/Mpc, and outperforms when $k>1$ h/Mpc. This is consistent with the fact that HOD is not being designed to work well on small scales. While the $P(k)$ of the benchmark method has significantly differ from the target's at $k=1$h/Mpc, $P(k)$ from our two-phase approach with Inception+R2Unet only begins to differ from the target at k $\approx$ 8 h/Mpc. 
This is likely due to the fact that the field is highly non-linear at those scales, which makes the model harder to learn. Furthermore, the fact that the galaxies are approximately poisson distributed produce a `shot-noise' floor. This affects the power spectra prominently on small scales: it adds a white component, $1/\bar{n}_{gal}$, to the power spectra, where $\bar{n}_{gal}$ is the average number density of the galaxies in the simulation box. Its effect can be clearly seen in Fig. ~\ref{fig:power}.

\subsection{Three-Point Correlation and Bispectrum}
The Universe, on large-scales, resembles a Gaussian field, and therefore, can be fully characterized by its 2pt correlation function or power spectrum. However, on small scales, non-linear gravitational evolution changes the density field into a non-Gaussian field. In order to characterize the spatial distribution of the Universe on small scales, where most of the cosmological information lies, higher-order statistics are needed. Here we concentrate on the bispectrum, the Fourier equivalent of 3 point correlation function, defined as

\begin{equation}
    B(k_1,k_2,k_3) \delta(\bm k_{123}) = \langle \delta_{\bm k_1} \delta_{\bm k_2} \delta_{\bm k_3} \rangle
\end{equation}

\begin{figure}
\vspace{-1cm}
  \centering
    \includegraphics[width=0.70\linewidth]{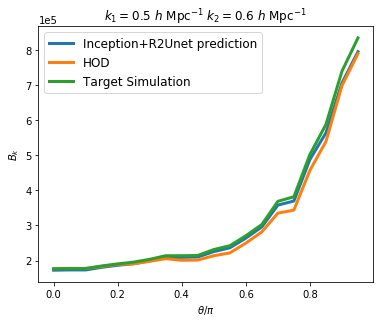}
    \includegraphics[width=0.69\linewidth]{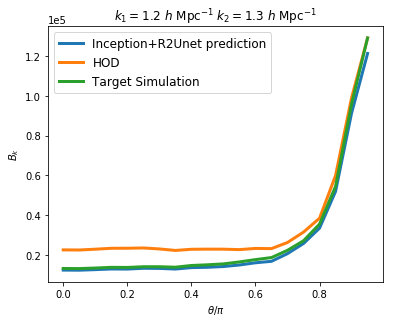}
    \caption{Bispectrum comparison between our Inception+R2Unet model, benchmark HOD model and Target Simulation. Upper panel: Bispectrum at $k_1=0.5$ $h$/Mpc and $k_2=0.6$ $h$/Mpc ("large" scale). Bottom Panel: Bispectrum at $k_1=1.2$ $h$/Mpc and $k_2=1.3$ $h$/Mpc ("small" scale). Inception+R2UNet outperfoms benchmark HOD model on both scales, and significantly so on smaller scales. }

    \label{fig:bispec}
\end{figure}

Figure \ref{fig:bispec} shows the bispectrum for our Inception+R2Unet model, the benchmark HOD model and the target.
At large scales, small wavenumber (k), the bispectra of our model and benchmark models are consistent with that of the target. The mean relative bispectrum residual of our model and HOD model compared to the target at $k_1=0.5$ h/Mpc and $k_2=0.6$ h/Mpc is 2.7\% and 5.0\% respectively.
On smaller scales, at $k_1=1.2$ h/Mpc and $k_2=1.3$ h/Mpc, the corresponding mean relative residual is 0.68\% and 1193\%. Our model reproduces the highly non-linear galaxy field on small scales far better than the benchmark model. This suggests that our machine learning model has large enough flexibility to reproduce the galaxy distribution from large to small scales even when we consider the higher order function, while the state-of-art benchmark produces bispectrum that is 5-6 times larger than the target at all scales.

\subsection{Voids}
The so-called cosmic web, i.e. the spatial distribution of matter on the Universe, is made up of high-density regions, clusters, where hundreds or thousands of galaxies resides. Galaxy clusters are connected by medium-density regions that contain highly ionized gas; the filaments. Finally, filaments are surrounded by enormous empty regions named voids. These voids are a very important element of the cosmic web, since most of the volume of the Universe resides on them. Given their unique nature, they embed a large amount of cosmological information. In this work we study the abundance of voids, as a function of their radii: the void size function. We identify voids in the galaxy distribution of the different models and the target using the algorithm described in \cite{Arka_16}.

\begin{figure}
\vspace{-1cm}
  \centering
    \includegraphics[width=0.70\linewidth]{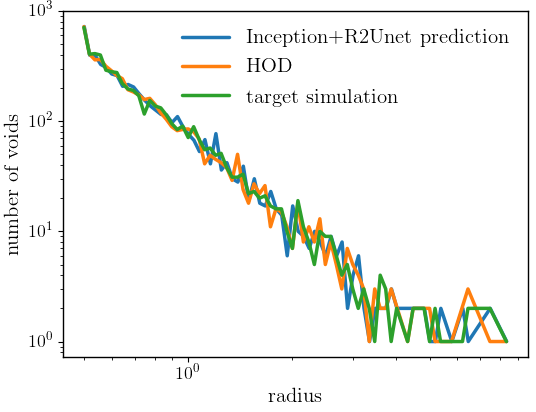}
    \caption{Number of voids as a function of their radii for the HOD (orange), Inception+R2UNet (blue) and the target (green). Both models are able to reproduce the abundance of voids from the target with high accuracy.}
    \label{fig:voids}
\end{figure}

We compare the void size function from our Inception+R2Unet model, the benchmark HOD model and the target simulation. The results are presented in Figure \ref{fig:voids}. The size function of voids from Inception+R2Unet model and benchmark HOD are both consistent with that of the target simulation, indicating our R2Unet+Inception model is competitive against the benchmark in this large scale observable.  

\subsection{Training time}

Another key aspect of our approach is its scaling abilities. 
Table \ref{tab:runningtime} shows the time needed to train and/or generate one simulation of galaxies using these various methods. HOD takes comparable amount of time to optimize. However, once trained, our method can generate large volumes of galaxy distribution with negligible time, and is flexible enough to generate excellent match to the target from small to large scales.

\begin{table}
\centering
\caption{Running time of different models}
\begin{tabular}{p{3cm}lll}
\toprule  
Model                        & Device            & CPU/GPU Hours \\ \midrule
Illustris        & CPU           & 19 million      \\
HOD        & 1CPU    &       8     \\
Inception + R2Unet        & 1GPU (GTX1080)          & 3      \\

\bottomrule
\end{tabular}
\label{tab:runningtime}
\end{table}

\section{Conclusions}

In this paper, we present a deep-learning based approach to model the link between the underlying dark matter from N-body simulations and the galaxies distribution from full hydrodynamic simulations. We design a specific learning scheme and model architecture to overcome the very high sparsity of the task. We show that our approach, by optimizing directly the number of galaxies prediction per voxel, manages to reproduce a large variety of important physical properties of the original data. It outperforms, or is as efficient as the standard benchmark method of the field, on various important cosmological statistics, while having much more scaling and generalization abilities. This is a first encouraging step to overcome the need for computationally expensive hydrodynamic simulations in the long run.

This work opens several trails for future research.
First, it will be very interesting to extend our model to be able to predict not only the number and positions of the galaxies but also their internal properties, e.g. stellar mass, star-formation rate, metallicity, etc. Training the model at different epochs of the Universe will allow us to better understand the complicated physics involved in galaxy formation/evolution. Training our model on simulations with different strengths of the most relevant astrophysical processes, such as active galactic nuclei and supernova feedback, will enable us to marginalize over these astrophysical complications and extract robust cosmological information.

Our approach can be used to populate the dark matter halos of very big gravity-only simulations with galaxies, without relying on the standard assumptions involved in the classical HODs. This will open new doors in cosmology, allowing us to investigate some of the most important theoretical systematics on cosmology such as baryonic effects. Since our framework allow us to model the spatial distribution of realistic galaxies down to very small scales, it can be used to extract the maximum information from cosmological observations. Our results can thus have a major impact on cosmology and will establish a new link between astrophysics and cosmology.

\section*{Acknowledgement}
We thank David Spergel, Siamak Ravanbakhsh and Barnabas Poczos for insightful discussions.  This project is supported by Center for Computational Astrophysics of the Flatiron Institute in New York City. The Flatiron Institute is supported by the Simons Foundation.

\bibliographystyle{ACM-Reference-Format}
\bibliography{example_paper}

\newpage
\clearpage
\section*{Supplement}
\subsection*{Data Access and Preparation}
The data used for this work is publicly available at Illustris website\footnote{\url{http://www.illustris-project.org/data/}}. 

We use Illustris-1 (hydrodynamical simulation) and Illustris-1-Dark (N-body Dark matter particle only simulation), which have the largest resolutions.
The simulation consists of 6,028,568,000 dark matter particles and of an equal number of hydrodynamic voronoi cells in the hydrodynamic simulation. 

We use the snapshot at redshift $z=0$, which is the current universe. It has a volume of $75^3 ({\rm Mpc/h})^3$ (\textit{Megaparsec}, 1 Mpc = 3.09$\times$ $10^{22}$ meters). The governing cosmological parameters are $\Omega_m$ = 0.2726, $\Omega_{\Lambda}$=0.7274, $\Omega_b$=0.0456, $\sigma_8$ = 0.809, $n_s$ = 0.963, $H_0$ = 100 $h$ $km s^{-1} Mpc^{-1}$ and $h$ = 0.704.

We take the positions of dark matter particles (from Illustris-1-Dark) and subhalos (from Illustris-1, where galaxies reside in), and grid the box into $1024 \times 1024 \times 1024$ pixels, obtaining 32786 non-overlapping sub-cubes of size $32 \times 32 \times 32$. We count the number of particles/subhalos in each pixel using Nearest Gird Point method. We use the top-left $13 \times 13 \times 31$ subboxes for validation, the bottom right $18 \times 18 \times 18$ subboxes for test, and the rest of subboxes are used for training.

\subsection*{Hyper-parameters search and Model Configuration}
 
Table.~\ref{tab:config} shows the various hyper-parameters and the search space for each parameter.

\begin{table}[h]
\centering
\caption{Hyper-parameters space search}
\resizebox{\columnwidth}{!}{%
\begin{tabular}{p{3cm}lll}
\toprule  
Configurations                        & Description & Search Space\\ \midrule

lr        & learning rate   & $10^{-5} - 10^{-3}$  \\
epochs &number of epochs&20-40\\ 
batch\_size &batch size&16, 32\\
loss\_weight &weight $w$ of the loss function&0.6-80\\
conv1\_out &number of hidden units for the 1x1x1  kernel&3-40\\
conv3\_out &number of hidden units for the 3x3x3 kernel&4-60\\
conv5\_out &number of hidden units for the 5x5x5 kernel&5-80\\
optimizer & optimizer for training & SGD, Adam\\
\bottomrule
\end{tabular}
}
\label{tab:config}
\end{table}
The best configuration for the classifier (the first phase) in the two-phase model (Inception + R2Unet) is: batch\_size=16, conv1\_out=6, conv3\_out=8, conv5\_out=10, epochs=20, loss\_weight=80, lr=0.001, optimizer = Adam.

And the best configuration for the regression (the second phase) in the two-phase model (Inception + R2Unet) is: batch\_size=16, epochs=20, loss\_weight=0.8, lr=0.001, optimizer = Adam.

\subsection*{Training Setup}
The experiments are carried out on NYU HPC cluster with one Intel Xeon E5-2690v4 2.6GHz CPU, one NVIDIA GTX 1080 GPU and 60GB of RAM.  We use Python version 3.5.3 and Pytorch version 0.4.1. The code is available in Github Repository: \url{https://github.com/xz2139/From-Dark-Matter-to-Galaxies-with-Convolutional-Networks}. The codes for evaluation on HOD (power spectrum, bispectrum and void finder) are also available publicly at the following Github Repository \url{https://github.com/franciscovillaescusa/Pylians}.
\\~\\
As we described in Section 4, our model has two phases: classifier (first phase) and regression (second phase). In the first classifier phase, we train a classifier to get the binary prediction for the location of the galaxies (Algorithm.~\ref{classifier}). Since the output of the model is the binary prediction, we convert target density fields into binary values before training the model (Algorithm \ref{binary_target}). 
 
\begin{algorithm}[h]
 \KwData{All Training Sub-cubes (Targets Only)}
 \KwResult{Binary Targets}
 \For{i in all training range(sub-cubes)}{
  \eIf{Targets[i] > 0}{
  Binary\_Targets[i] = 1\;
  }{
  Binary\_Targets[i] = 0\;
  }
 }
 \caption{Turn targets into binary value}
 \label{binary_target}
\end{algorithm}

\begin{algorithm}[h]
 \KwData{All Training Sub-cubes (Inputs and Binary\_Targets)}
 \KwResult{Binary\_Prediction}
 \For{i in all training range(sub-cubes)}{
  Binary\_Prediction = Inception\_Net(sub-cubes[i])\; 
  Loss = Cross\_Entropy(Prediction, Binary\_Targets[i])\;
  Back-propagation\;

 }
 \caption{Classifier Running Process}
 \label{classifier}
\end{algorithm}

The binary prediction from the first phase will then serve as a mask for the second phase, where only the masked region is trained on. 
In the second phase regression model, 
the input of the model is the same as the classifier model, but the targets are the real number of galaxies. 

To achieve the masking, we multiply the outputs from the regression model with our binary prediction from the classifier model. 
The algorithm for the second phase is shown in Algorithm \ref{final_step}.
\begin{algorithm}[h]
 \KwData{All Training Sub-cubes (Inputs and Real\_Targets), Binary\_Prediction from the first phase}
 \KwResult{Final\_Prediction}
 \For{i in all training range(sub-cubes)}{
  Prediction = R2UNet(sub-cubes[i])\; 
  Final\_Prediction = Binary\_Prediction $\cdot$ Prediction\;
  Loss = L2\_Loss(Final\_Prediction, Real\_Targets[i])\;
  Back-propagation\;
 }
 \caption{Regression Model Running Process}
 \label{final_step}
\end{algorithm}

%
\end{document}